\documentclass[9pt,twocolumn,twoside]{opticajnl}
\journal{opticajournal} 
\setboolean{shortarticle}{true}

\usepackage{lineno}
\usepackage{braket}
\usepackage{adjustbox}
\usepackage{ulem} 

\title{Supercontinuum generation from Topological Edge Supermodes in a short SSH Photonic Crystal Fiber}

\author[1]{Daniel Rodriguez-Guillen}
\author[1]{Carlos Wiechers}
\author[1,*]{Lorena Velazquez-Ibarra}

\affil[1]{Departamento de F\'isica, Divisi\a'on de Ciencias e Ingenier\a'{\i}as, Universidad de Guanajuato, 37150, Le\'on, Guanajuato, M\'exico}
\affil[*]{lorenav@fisica.ugto.mx}

\begin{abstract}
We introduce a topological photonic-crystal fiber that embeds a short Su–Schrieffer–Heeger (SSH) chain and supports two edge supermodes. Using full-vector modal analysis and a coupled generalized nonlinear Schrödinger equation, we show that each supermode provides an independent nonlinear channel with a distinct broadening mechanism: the even supermode features two zero-dispersion wavelengths and yields degenerate four-wave mixing sidebands, whereas the odd supermode is all-normal-dispersion and generates a smooth, flat ANDi-type continuum. Exciting a single core prepares a coherent superposition of the two supermodes; cross-phase modulation and inter-parity four-wave mixing then enable energy transfer across detunings inaccessible to either mode alone, producing the broadest and flattest spectrum with new short wavelengths components. Our results establish topology-enabled modal control as a scalable knob for engineering supercontinuum generation in short SSH topological fibers.
\end{abstract}

\setboolean{displaycopyright}{false} 

\begin{document}

\maketitle
\begin{figure}[H]
  \centering
  \includegraphics[width=.80\columnwidth]{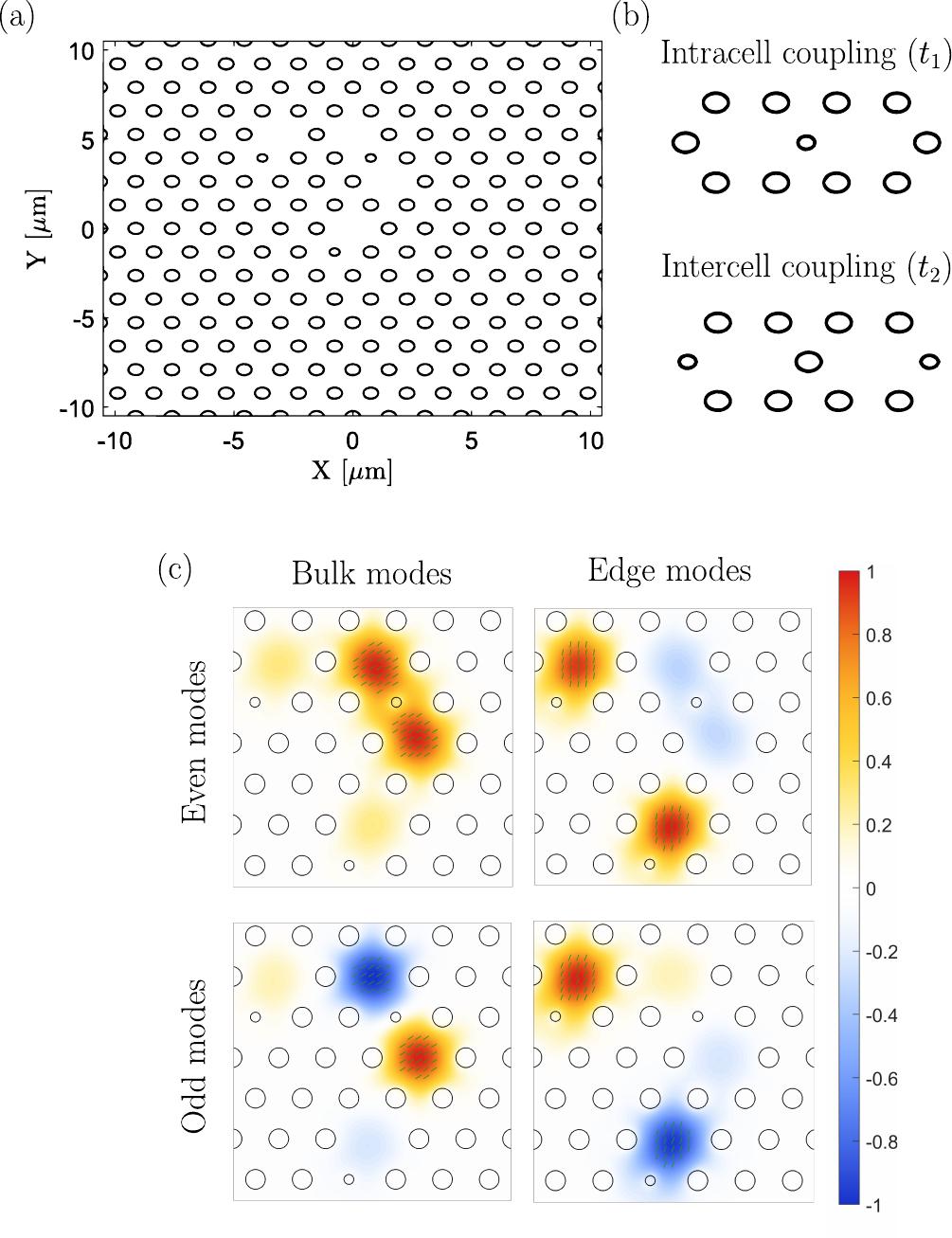}
  \caption{(a) Cross-section of the PCF embedding a four-core SSH chain.
  (b) SSH sketch with intracell coupling $t_{1}$ and intercell coupling $t_{2}$; edge
  states occur for $t_{1}<t_{2}$. (c) Representative normalized field maps (color:
  $|E|/\max|E|$) for bulk (left) and edge (right) supermodes in the even (top) and odd
  (bottom) parity sectors. The even/odd edge states are the hybridized end modes used
  throughout. All panels use the same spatial scale; maps shown at the pump wavelength
  $\lambda_{0}=1.064~\mu\mathrm{m}$ and one polarization.}
  \label{fig:1}
\end{figure}

Supercontinuum generation (SCG) in optical fibers carries applications ranging from precision frequency metrology and comb spectroscopy to advanced microscopy, and is enabled by photonic crystal fibers (PCFs) with highly tailorable dispersion~\cite{Dudley06, Udem02, Picqué19, ji15}. The underlying physics (soliton dynamics, dispersive waves, all-normal-dispersion broadening, Raman contribution, and shock terms) is well established, with robust generalized nonlinear Schrödinger-equation (GNLSE) models~\cite{Laegsgaard07}. In parallel, topological photonics~\cite{ozawa19}, has emerged as a framework for robust light transport via bulk–boundary correspondence; the Su-Schrieffer-Heeger (SSH) model with alternating couplings $t_1$ (intracell) and $t_2$ (intercell) provides a canonical platform for photonic edge states~\cite{Roberts22, Bergamasco19, Michelle19, Mittal18, Kang23, Noh18, Hyok23}. 
Figure~\ref{fig:1} summarizes the concept and design: Fig.~\ref{fig:1}(a) shows the design of the PCF cross-section embedding an SSH four-core chain; Fig.~\ref{fig:1}(b) the SSH sketch with $t_1$, $t_2$.; and Fig.~\ref{fig:1}(c) representative even/odd edge modes versus bulk modes.

\begin{figure}[H]
  \centering
  \includegraphics[width=.88\columnwidth]{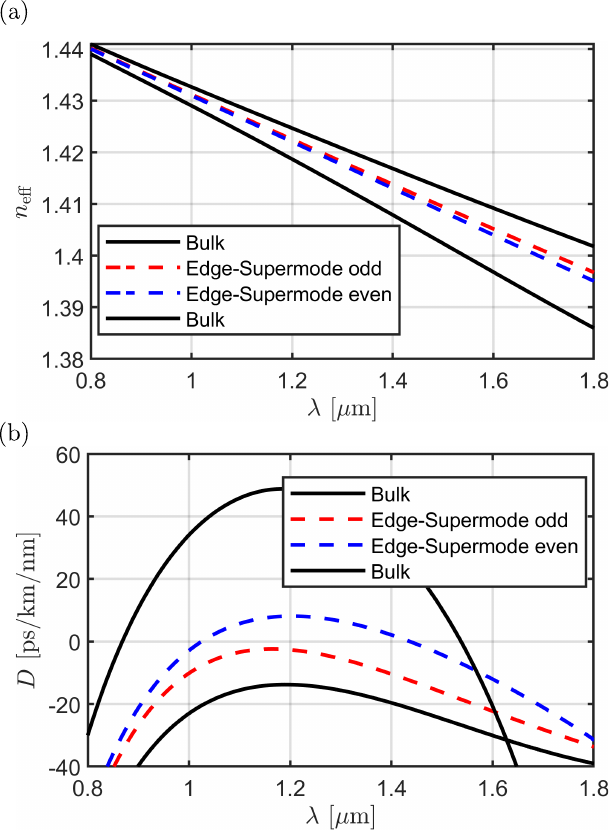}
  \caption{(a) Effective index, $n_\mathrm{eff}(\lambda)$, for bulk modes (black solid) and edge supermodes (odd, red dashed; even, blue dashed). (b) Corresponding dispersion, $D(\lambda)$. The even and odd edge supermodes exhibit distinct dispersion profiles, so each provides a different nonlinear channel for SCG.}
  \label{fig:2}
\end{figure}

In a short finite even-length SSH chain, two edge states hybridize into parity-defined edge supermodes (even/odd) with propagation constants $\beta_e(\omega)$/$\beta_o(\omega)$  (Fig.~\ref{fig:1}(c)). In the supermode basis, propagation does not exhibit coupling between supermodes, whereas in the core basis the core excitation beats with a coupling length $L_c = \pi/(\beta_e - \beta_o)$. 
Because $\beta_{e}(\omega)$ and $\beta_{o}(\omega)$ differ, their effective index, $n_{eff}(\lambda) = \beta/k_0$, and chromatic dispersion, $D(\lambda)$, differ as well; thus each edge supermode is an independent nonlinear channel. Figure~\ref{fig:2}(a) shows the effective refractive index and Fig.~\ref{fig:2}(b) the chromatic dispersion for each edge and bulk branches, highlighting the even/odd dispersion split to obtain two SCG mechanisms.

To simulate SCG in the supermode basis, we use full-vector finite-difference frequency-domain (FDFD) mode solving to calculate the effective areas and nonlinear coefficients used in the multimode coupled GNLSE \cite{Stajanca16, Poletti08, Biaggio14, su22}. From the transverse fields of the even/odd edge supermodes we compute the effective area, $A_{eff}^{e,o}(\lambda)$, and the four Kerr coefficients defined by the mode-overlap integrals: $\gamma_{eeee}(\lambda)$ and $\gamma_{oooo}(\lambda)$ (Self-phase modulation (SPM) for even/odd), $\gamma_{\text{xpm}}(\lambda) = \gamma_{eeoo} = \gamma_{ooee}$ (Cross-phase modulation (XPM) between even/odd), and $\gamma_{\text{fwm}}(\lambda) = \gamma_{eooe} = \gamma_{oeeo}$ (inter-parity phase-conjugate four-wave mixing (FWM)). These, together with $\beta_{e,o}(\omega)$, fully specify the coupled model; driving a single supermode reduces to the standard GNLSE. Supplementary material shows all effective areas with the FWM/XPM overlaps, and the nonlinear parameters. 

The topological PCF is designed on a hexagonal lattice compatible with the stack-and-draw fabrication technique. To realize the SSH configuration, we choose the shortest even chain (four cores) and tune the intracell link by shrinking the intracell air hole; the air-hole ratio $r_1/r_2=2$ (cladding-lattice radius $r_1$ intracell radius $r_2$) strengthens intracell coupling and improves long-wavelength confinement, with $\Lambda=1.52~\mu$m and intercell/intracell radii of $0.320$/$0.160~\mu$m. 
Full-vector simulations use a square computational window of $23~\mu$m with spatial step $\Delta x=0.021~\mu$m, selected according to the rule of thumb $\Delta x=\lambda_{\min}/(n_{\max}\times\mathrm{ppw})$, taking $\lambda_{\min}=0.4~\mu$m, $\mathrm{ppw}=13$, and $n_{\max}=1.45$. To improve accuracy, we implemented subpixel smoothing~\cite{Alexopoulos:22} and a perfectly matched layer (PML), achieving precision consistent with our previously validated tests~\cite{Yo25}. 

We swept the wavelength from $0.4$ to $1.8~\mu$m, and inferred the SSH couplings from the four propagation constants (two bulk, two edge) without fitting via analytic four-site relations.
\begin{figure}[t]
  \centering
  \includegraphics[width=.88\columnwidth]{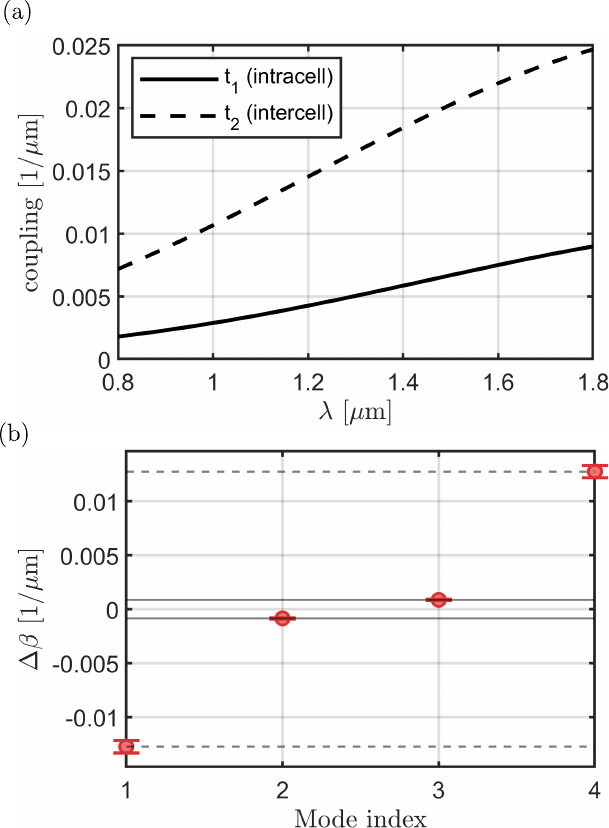}
  \caption{(a) Coupling coefficients extracted from the four-mode spectra: intracell $t_1(\lambda)$ (solid) and intercell $t_2(\lambda)$ (dashed). Across the band $t_2>t_1$, indicating the non-tivial SSH phase. (b) Robustness at the pump wavelength $\lambda_0 = 1.064~\mu$m, eigenvalue detunings $\Delta\beta$ (relative to the mid-gap $\beta_0$) under off-diagonal coupling disorder with r.m.s. 0.05 max$\{(t_1,t_2)\}$ (over 1000 realizations). Red circles: mean; red error bars: $\pm1~\sigma$. Solid gray lines mark the clean edge detunings $\pm\zeta_1$; dashed gray lines mark the clean bulk detunings $\pm\zeta_2$. Error bars for the two edge modes are not appreciable at this scale, while bulk modes show a larger (still visible) variance.}
  \label{fig:4}
\end{figure}
We consider $\beta_o(\lambda)$ the mid-gap average of the two edge branches and define the absolute detunings of the edge and bulk branches as $\zeta_1(\lambda)$ and $\zeta_2(\lambda)$. The intracell and intercell couplings are defined as $t_1(\lambda) = \sqrt{\zeta_1(\lambda)\zeta_2(\lambda)}$, $t_2(\lambda) = \zeta_2(\lambda) -\zeta_1(\lambda)$. Equivalently, the supermode detunings $\Delta\beta$ and the eigenvectors $\ket{\mathbf u}$ satisfy the coupled-mode eigenproblem $C\ket{\mathbf u}=\Delta\beta\ket{\mathbf u}$~\cite{Roberts22, Bergamasco19}, where $C$ is the coupling matrix, so the calculated propagation constants $\beta$ uniquely determine $t_{1,2}$ for each wavelength. Figure~\ref{fig:4}(a) shows $t_2>t_1$ across the band, i.e. the non-trivial SSH phase. Robustness is assessed by introducing coupling (off-diagonal) disorder that preserves the SSH chiral symmetry: each coupling is modified as $t_j \xrightarrow{} t_j(1+\delta_j)$ with $\delta_j \sim \mathcal{N}(0,\sigma_c^2)$, $\sigma_c = 0.05$. We then diagonalize the disordered coupling matrix $C$ for $10^3$ realizations. Fig.~\ref{fig:4}(b) shows the eigenvalues $\Delta\beta$ (mean $\pm$ s.d.) of the two edge modes and the two bulk modes for the pump wavelength $\lambda_0 = 1.064$~$\mu$m. The two edge modes remain isolated with a narrow spread (on average $6.8\times10^{-5} $~$\mu$m$^{-1}$), whereas the bulk modes shift more strongly (on average $5.7\times10^{-4} $~$\mu$m$^{-1}$). Their mid-gap positions are not exactly zero because a finite chain hybridizes the two edge states, resulting in a splitting, with additional bias from weak chiral-symmetry breaking due to the material dispersion and non-identical overlaps. These couplings, together with $\beta_{e,o}(\omega)$ and the nonlinear coefficients, are used on the coupled GNLSE solved by a fourth-order Runge–Kutta in the interaction picture (RK4IP) with adaptive step, including Raman contribution and shock term~\cite{Hult07, Heidt09}.

\begin{figure}[t]
  \centering  \includegraphics[width=\columnwidth]{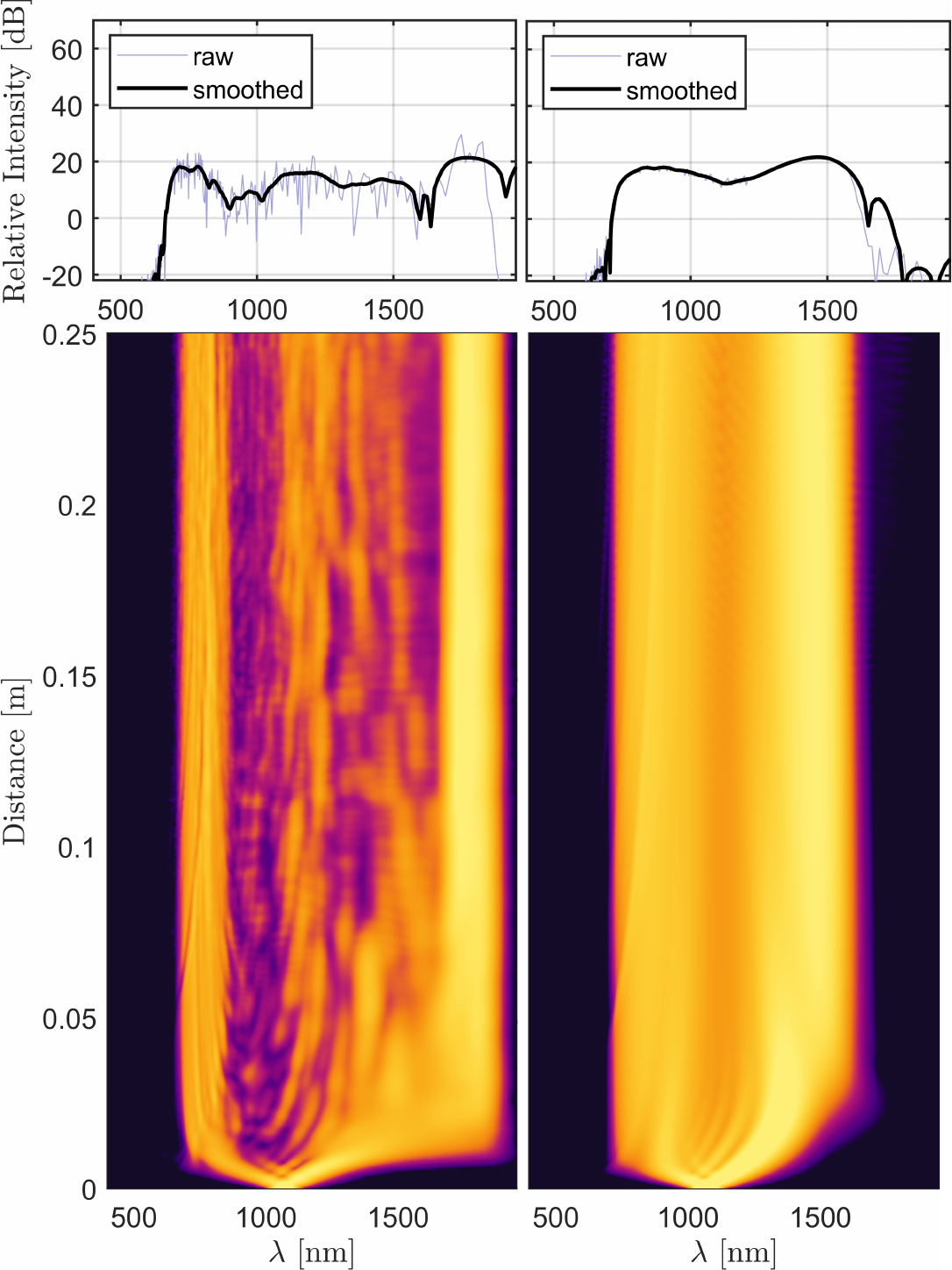}
  \caption{SCG for eigenmode launches. Left: even edge supermode showing FWM sidebands and subsequent cascades. Right: odd edge supermode showing smooth SPM-dominated broadening. Top: output spectra (raw/thin and smoothed/bold); bottom: evolution maps (yellow-to-purple colormap).}
  \label{fig:5}
\end{figure}
\begin{figure}[t]
  \centering
  \includegraphics[width=\columnwidth]{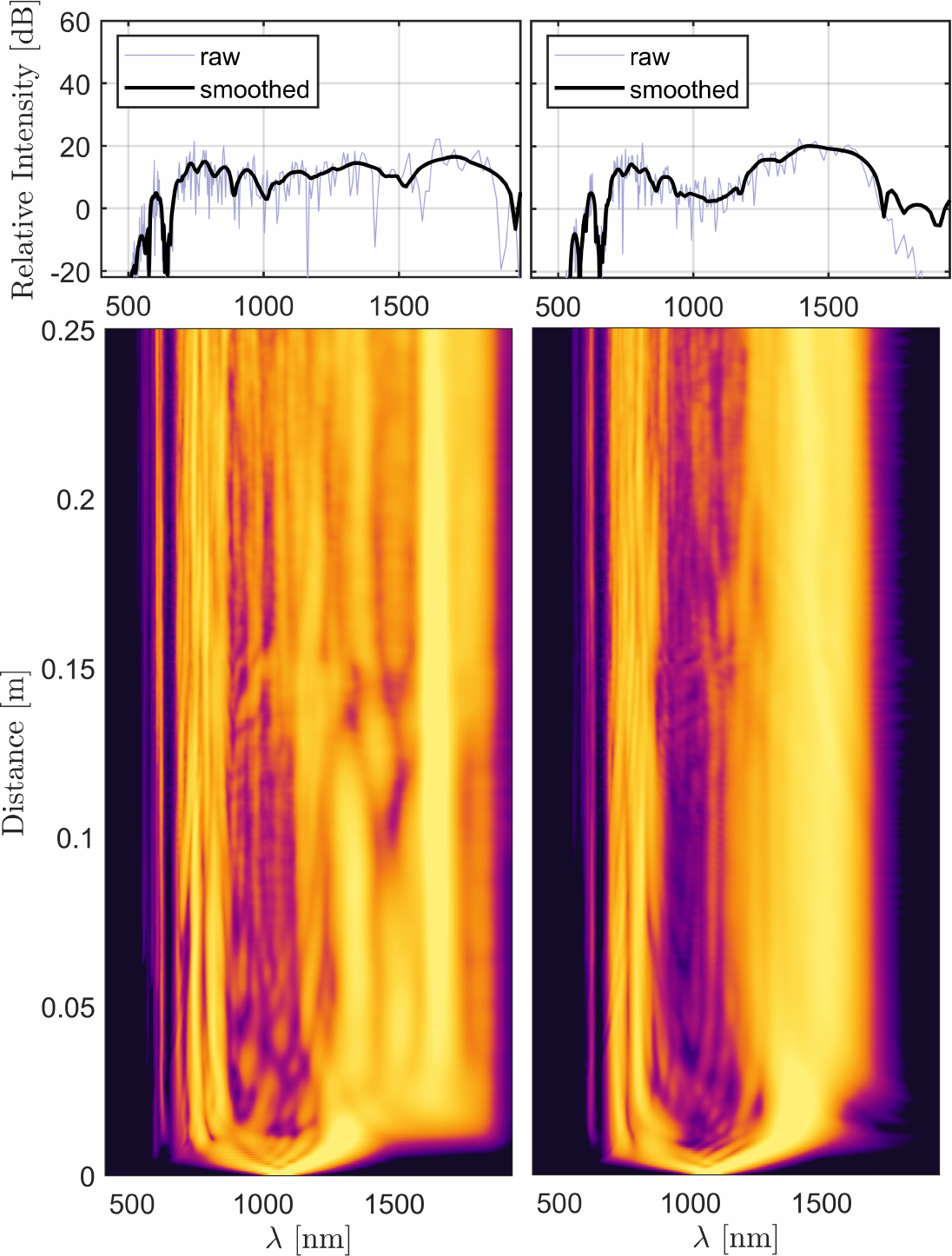}
  \caption{SCG for single-core launch, analyzed in the even (left panel)/odd (right panel) supermode basis. Both projected spectra inherit the hallmarks of their eigenmode analogs but extend further due to XPM and inter-parity FWM enabled by coherent co-propagation. Top: output spectra (raw/thin and smoothed/bold); bottom: evolution maps (yellow-to-purple colormap).}
  \label{fig:6}
\end{figure}
Supercontinuum generation was simulated in three standard launching conditions: (i) Even (symmetric) edge supermode excitation: because the even supermode exhibits two zero-dispersion wavelengths (ZDWs), degenerate four-wave-mixing phase matching occurs within this band and seeds two sidebands; (ii) Odd (antisymmetric) edge supermode excitation: since the dispersion is completely normal, a broad and flat ANDi-type supercontinuum is obtained~\cite{Bres23}; (iii) Single-core excitation: launching light into a single core excites a fixed superposition of even and odd edge supermodes determined by modal overlaps; in the supermode basis the dynamics are described by two coupled GNLSEs with self- and cross-phase modulation (and inter-mode FWM if phase-matched). All SCG spectra are obtained by solving the coupled GNLSE with pump wavelength $1.064~\mu$m, pulse duration of $10$~fs, and pulse energy of $1.5$~nJ. We set the Raman fraction $f_R=0.17$, which contributes to the long-wavelength extension of the supercontinuum due to Raman energy transfer. 

Launching the even edge supermode (Figure~\ref{fig:5}, left panel) produces a spectrum structured by the two ZDWs; as a consequence, degenerate FWM is phase matched inside this band, SPM first broadens the pump, then FWM seeds symmetric sidebands that are mixed with dispersive waves. By contrast, launching the odd edge supermode leads to broadening governed by SPM, optical wave-breaking, and self-steepening. The resulting spectrum is smooth and comparatively flat without FWM bands (Figure~\ref{fig:5}, right panel).

A single-core pump excites a fixed coherent superposition of the even and odd edge supermodes. When we project the simulated field back onto the even and odd supermode bases, both share features of their eigenmode counterparts, and each spectrum becomes broader. The reason behind this is that the two channels co-propagate and exchange energy through XPM and inter-parity FWM, opening detunings inaccessible to either channel alone. Figure~\ref{fig:6} shows both the even projection (left panel) and the odd projection (right panel). The even supermode preserves sidebands and exhibits a flatter plateau between them, while the odd projection shows stronger intense bands. Both spectra show new bands, from $\sim530$~nm to $\sim630$~nm, obtained for an inter-parity FWM, generating more accessible wavelengths.
\begin{figure}[H]
  \centering
  \includegraphics[width=\columnwidth]{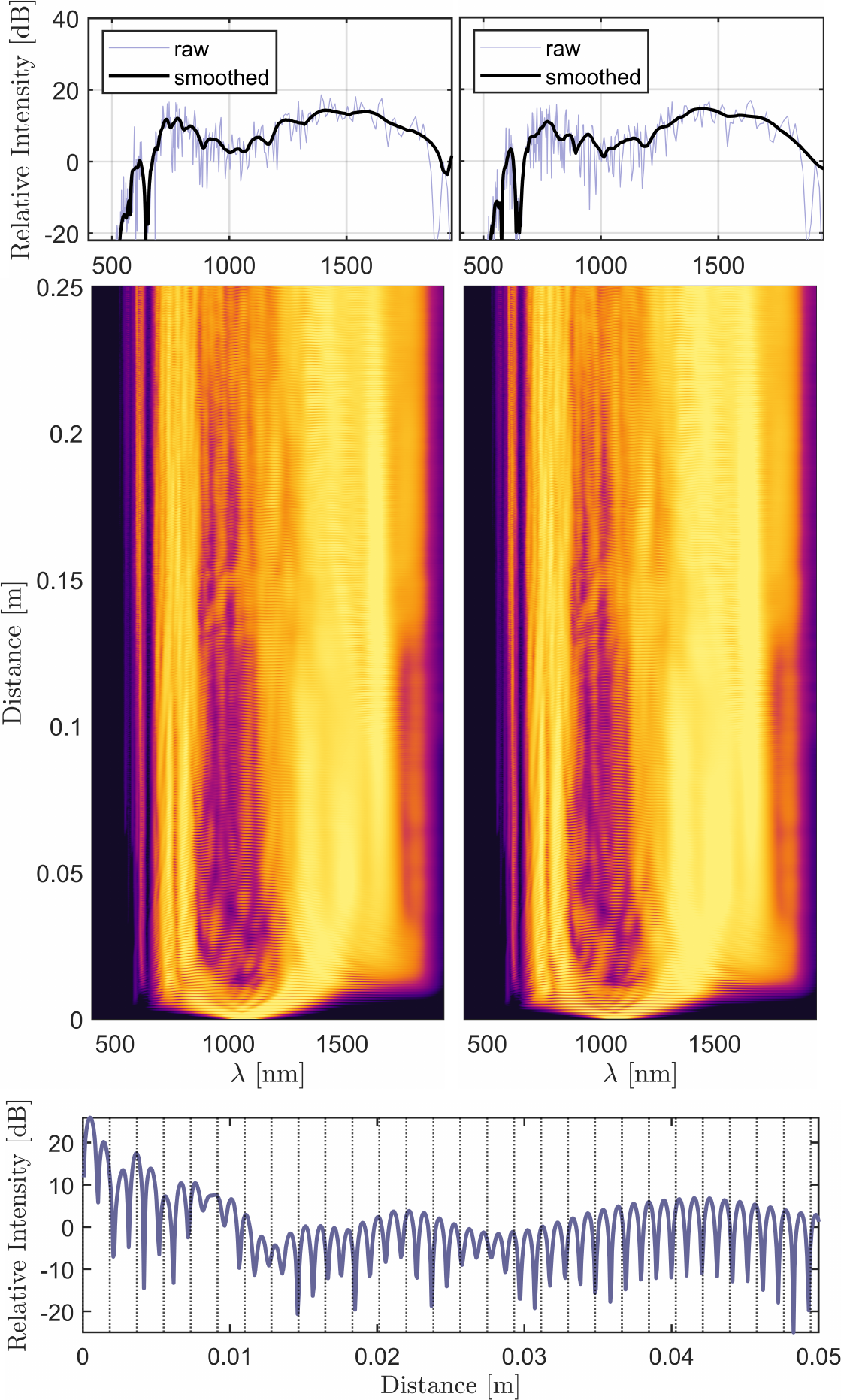}
  \caption{Core-resolved SCG. Top: right- and left-core spectra (raw/smoothed) and evolution maps; both are broad and similar because each is a fixed linear combination of the even/odd channels. Bottom: $1.064$~$\mu$m intensity vs. distance in the left core (blue) with the ideal single-wavelength beating envelope (black dashed) based on the coupling length.}
  \label{fig:7}
\end{figure}
Projecting onto each outer edge core (left and right cores of the SSH chain), produces two spectra that are very similar but not identical. They are similar because both projections are fixed linear combinations of the same even/odd supermodes; hence, a broader structure (SPM, FWM, and shock) appears in both. Small differences arise from the unequal projection of the effective area, and the phase differences between even/odd supermodes. Figure~\ref{fig:7} shows the two edge-core outputs (left edge core in left panel, and right edge core in right panel); both are broad and nearly flat, with a blue extension from inter-parity FWM and bands from degenerate FWM. Both spectra also show oscillations in intensity across distance (observable at the beginning of the propagation). These oscillations naturally occur due to the coupling parameter between edge supermodes modes. The bottom panel shows a 1.064~$\mu$m slice of the left-core intensity over the first 5~cm of the fiber, showing the oscillating nature; the dotted vertical lines are the coupling length ($L_c=\pi/\Delta\beta$). The simulated trace oscillates faster than the single-frequency envelope because the field is polychromatic and the channels nonlinearly exchange power (SPM/XPM/FWM), which modulates the beating period. 

In conclusion, we have introduced a topological photonic crystal fiber that embeds a short, even-length SSH chain and supports two edge supermodes with contrasting dispersion. By tuning the pitch and hole radii, the even supermode exhibits two zero-dispersion wavelengths, while the odd supermode operates in all-normal dispersion; each supermode acts as an independent nonlinear channel. We have numerically confirmed the non-trivial phase, as well as robustness to coupling disorder. Coupled GNLSE simulations show that the single-core pumping generates a coherent superposition of the two supermodes, due to the inter-parity FWM the spectrum becomes broader and flatter, which is not possible for each channel alone. Projection onto the edge-core basis shows coherent power beating modulated by a polychromatic nonlinear exchange. These results establish a topology-enabled modal control as a new degree of freedom, where engineering the coupling strengths and the modal mixing will enable reconfigurable supercontinuum sources.

\begin{backmatter}
\bmsection{Funding} This work has been funded in part by Secretaría de Ciencia, Humanidades, Tecnología e Innovación (SECIHTI, Ref. CBF-2025-I-2037).

\bmsection{Acknowledgment} D.R.-G. acknowledges financial support from SECIHTI (Graduate Fellowship, CVU-1142156).

\bmsection{Disclosures} The authors declare no conflicts of interest.

\bmsection{Data availability} The data that support the findings of this study are available from the authors upon reasonable request.

\end{backmatter}

\bibliography{biblio}
\end{document}